\documentclass[twocolumn,preprintnumbers,amsmath,amssymb,prl,superscriptaddress]{revtex4-1}
\usepackage{graphicx}
\usepackage[utf8]{inputenc}
\usepackage{dcolumn}
\usepackage{bm}
\usepackage{color}

\begin{document}
\title{Fate of doped carriers in silver fluoride cuprate analogues}

\author{Subrahmanyam Bandaru} 
\affiliation{Center of New Technologies, University of Warsaw, Zwirki i Wigury 93, 02089, Warsaw, Poland} 
\author{Mariana Derzsi}
\affiliation{Center of New Technologies, University of Warsaw, Zwirki i Wigury 93, 02089, Warsaw, Poland} 
\affiliation{
Advanced Technologies Research Institute, Faculty of Materials Science and Technology in Trnava, Slovak University of Technology, J. Bottu 25, 917 24, Bratislava, Trnava, Slovakia}
\author{Adam Grzelak} 
\affiliation{Center of New Technologies, University of Warsaw, Zwirki i Wigury 93, 02089, Warsaw, Poland} 
\author{José Lorenzana}
\email[]{jose.lorenzana@cnr.it}
 \affiliation{Institute for
Complex Systems-CNR, and Physics
Department, University of Rome La Sapienza, I-00185 Rome, Italy.}
\author{Wojciech Grochala}
\affiliation{Center of New Technologies, University of Warsaw, Zwirki i Wigury 93, 02089, Warsaw, Poland}


\date{\today}

\begin{abstract}
AgF$_2$ is a correlated charge-transfer insulator with properties remarkably similar to insulating cuprates which have raised hope that it may lead to a new family of unconventional superconductors  upon doping. We use {\em ab initio} computations to study doping strategies leading to metallization. Because the upper Hubbard band is very narrow electron doping leads to undesired strongly self-trapped states (polarons).  For the hole-doped case, polaron tendency is stronger than for cuprates but still moderate enough to expect that heavily doped compounds may become metallic. 
Since the strong electron lattice coupling originates in the strong buckling we study also an hypothetically flat 
allotrope and show that it has excellent prospect to become metallic.   
We compare the AgF$_2$ behavior with that for the hole-doped conventional cuprate La$_2$CuO$_4$ and electron-doped Nd$_2$CuO$_4$. Our results show a clear path to achieve high temperature superconductivity in silver fluorides. 
\end{abstract}
\maketitle
\begin{figure*}[t] 
 \centering
 \includegraphics[width=18 cm]{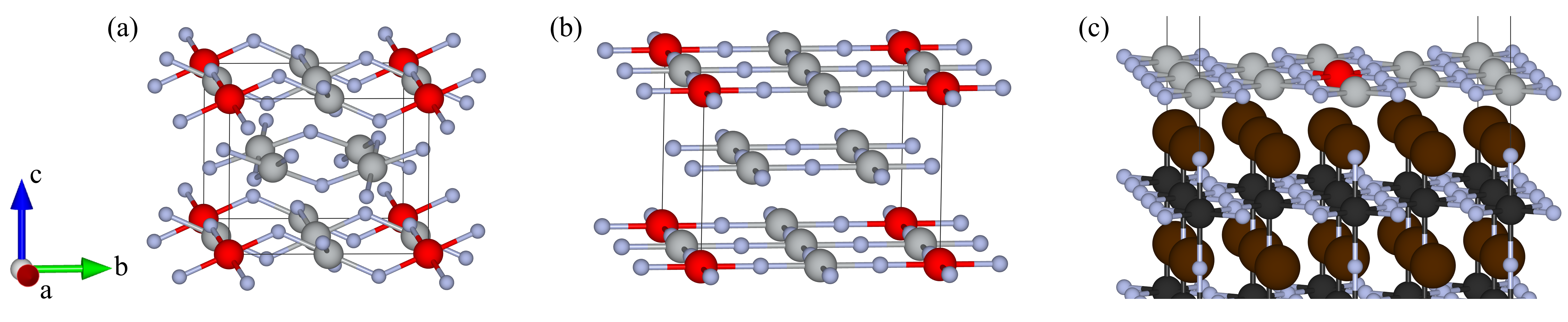} 
  \caption{AgF$_2$ structure.  We show $\alpha$-AgF$_2$ (a),  an hypotetical flat polymorph (b) and a monolayer of AgF$_2$ on 
a  RbMgF$_3$ substrate.   The thin lines show the supercells used for the search of one-polaron states. For (c), only a fragment of the unit cell,
is shown which comprises 7.5 layers of RbMgF$_3$ and 20 \AA~of vacuum region.
Atoms are  Ag (gray), F (light blue), Rb (dark brown), Mg (black).  
The polaron Ag site and equivalent sites are highlighted in red.}
  \label{fig:structure}
\end{figure*}

Since the discovery of high-$T_c$  superconductivity in cuprates by Bednorz and M\"uller\cite{Bednorz1986}
there have been many attempts\cite{Norman2016} to replicate the physics of these systems with different elements other than copper and oxygen.  One obvious direction is changing copper by silver, the next  coinage metal in group 11. However, the Ag 4$d^9$ levels are much deeper (with respect to vacuum) than Cu 3$d^9$ states. Thus AgO, {although} nominally  4$d^9$ is not magnetic and shows a negative charge transfer energy as realized\cite{Tjeng1990} in the early times of high-$T_c$ research.
 This drawback can be solved\cite{Grochala2001} by the substitution of O by the more electronegative F restoring a positive charge transfer energy and leading, at least in the insulating phase, to an excellent cuprate analogue\cite{Gawraczynski2019,Miller2020}. Commercially available 
AgF$_2$ is a correlated  charge transfer insulator with a superexchange interaction $J$ which is 70\% of a typical cuprate\cite{Gawraczynski2019}. The structure shown in Fig.~\ref{fig:structure}(a) appears as stacked AgF$_2$ planes, with the same topology as CuO$_2$ plane but much larger buckling. Another important difference is that a CuO$_2$ plane is not neutral and needs compensating ions while an AgF$_2$ plane {\em is} neutral thus AgF$_2$ is the simplest cuprate analogue of the silver fluoride family. Unlike cuprates, each ligand ion in AgF$_2$ has the double role of being planar and apical. Compared  to an hypothetical flat structure [Fig.~\ref{fig:structure}(b)] buckling allows for a (distorted) octahedra of F with a square planar coordination of Ag. Still, the inplane AgF bond\cite{Fischer1974} (2.07\AA) is substantially shorter than the apical bond (2.58 \AA) which leads to quasi two-dimensional\cite{Gawraczynski2019,Kurzydowski2018} magnetism.

The above premises set the stage to dope the material and search for unconventional high-$T_c$ superconductivity, an achievement that may
lead to a revolution in understanding and applications. The simplest approach would be to grow samples with fluorine non-stoichiometry, for example excess fluorine may lead to hole doped staged phases as excess oxygen\cite{Wells1997} in cuprates and recent proposals for Cu fluorides\cite{Rybin2020}.
 By the same token, fluorine deficiency may lead to electron doping. In cuprates excess oxygen has being achieved by electrochemical techniques and similar strategies are currently being explored in silver fluorides\cite{Poczynski2019}. Fluorine deficiency occurs spontaneously under illumination (which requires the use of small fluency in Raman\cite{Gawraczynski2019} experiments). Other possible doping strategies include field effect doping\cite{Grzelak2020} and partial substitution of Ag(II) by other transition metal cations\cite{Domanski2020}.  While the requirement of having a uniform dopant distribution could be extremely challenging, before embarking in this endeavor it would be useful to know what are the more promising strategies to achieve matalization in the first place.  

From another perspective, it has been proposed that monolayers of AgF$_2$ with various degrees of buckcling can be grown on appropriate substrates \cite{Grzelak2020}.
Bandwidth control by buckling has being a formidable tool to understand the physics of transition metal oxides\cite{Torrance1992,Imada1998}. Clearly, a cuprate analogue with a controllable buckling provides an excelent starting point for such investigations.  

One feature that would prevent metallization is the formation of strongly self-trapped polaronic\cite{Anisimov1992,Yonemitsu1992,Dobry1994,Lorenzana1994} states.
On one hand, the $4d$ orbitals of silver are expected to be more diffuse, thus less correlated than the $3d$ orbitals of copper favoring metallization. On the other hand, the strong buckling in the more common form of the compound (dubbed $\alpha$-AgF$_2$) reduces the bandwidth and increase the coupling with the lattice, both effects favoring polaron formation. While in cuprates the Zhang-Rice singlet\cite{Eskes1988,Zhang1988} (or its mean-field magnetic polaron version) couple with CuO stretching modes\cite{Anisimov1992,Yonemitsu1992,Dobry1994,Lorenzana1994}, for $\alpha$-AgF$_2$ {the coupling with the lattice involves} also the softer bending modes. 
Indeed for a straight Ag-F-Ag bond the coupling with the bending mode is quadratic in the displacement\cite{Devereaux1995a} while for a buckled structure the coupling becomes linear and thus more relevant.

In this work we use density functional theory computations to study the fate of a doped carrier in $\alpha$-AgF$_2$. In view of proposals to grow flat AgF$_2$\cite{Yang2014,Yang2015,Grzelak2020}  we also study a hypothetical flat AgF$_2$ case [Fig.~\ref{fig:structure}(b)] and a monolayer on an appropriate substrate  [Fig.~\ref{fig:structure}(c)] and compare them with similar computations on cuprates\cite{Anisimov1992}.

For bulk AgF$_2$, supercells with 8 Ag atoms and 16 F atoms have been chosen to investigate polaron formation. The calculations have been performed using the GGA + $U$ method with PBE\cite{Perdew1996} functional and  parameters close to accepted values in the literature, namely $U=5$ eV and $J=1$eV for AgF$_2$\cite{Gawraczynski2019,Kasinathan2007} and $U=9$eV and $J=1$ eV for cuprates\cite{Himmetoglu2011}. Sensitivity to $U$ is discussed below.  First, the undoped structures were optimized. In the case of puckered AgF$_2$ a full optimization was done while for hypothetical flat systems constraints were imposed to keep the Ag-F-Ag angle fixed at 180$^\circ$. Then, one electron or hole was added to the supercell (corresponding to a doping of $\pm1/8$) assuming a uniform compensating background. 
To induce a polaron a spin was flipped on the Ag site marked as red in Fig.~\ref{fig:structure}. As a first step, a calculation with fixed atomic positions was performed to stabilize a purely electronic magnetic polaron. Next, atomic positions were relaxed. All the calculations have been performed at a cut-off energy of 700 eV and using a high k-mesh of 8x8x8 with 256 k-points.

\begin{figure}[t] 
 \centering
\includegraphics[width=7.3 cm]{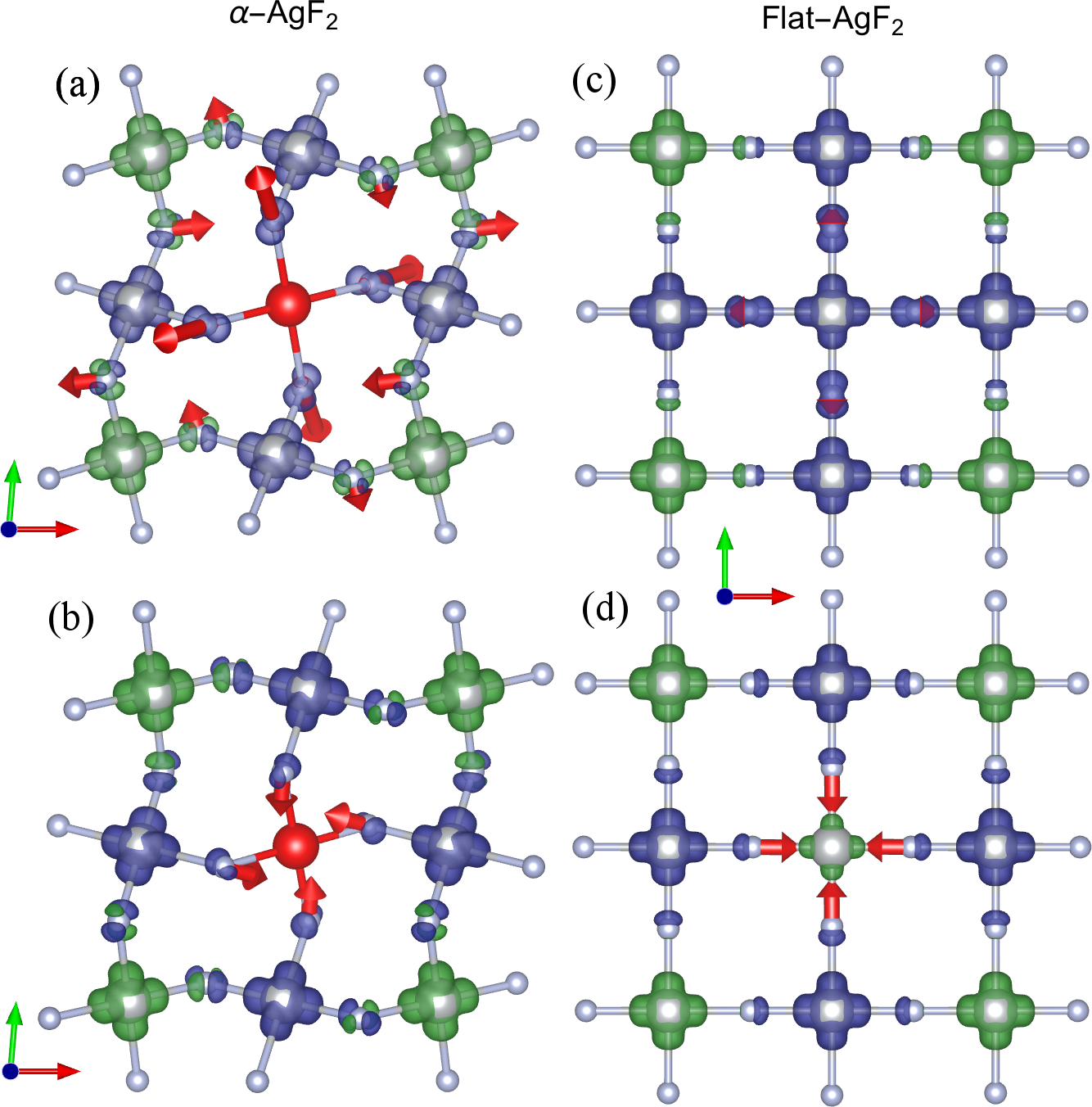} 
  \caption{Polaron solutions for $\alpha$-AgF$_2$ (first column) and flat-AgF$_2$ (second column). (a) and (c) are electron doped while (b) and  (d) are holes doped. 
The ball and stick model show the lattice relaxed solution. 
 The red arrows show the atomic displacements with respect to the undoped solution multiplied by a factor of 10 (a) and (b) or 20 (c) and (d). We show positive/negative magnetization isosurfaces in green and blue. In the flat electron doped case (c) the polaron solution is not stable so the resulting solution is actually a delocalized electron with a spin-flip at the center. 
}
  \label{fig:polaron}
\end{figure}

The initial spin-flip on the undoped solution produces two in-gap states one which is empty coming from the upper Hubbard band and one full from the fluorine valence band. Those  in-gap states host the added electron or hole and self-consistently stabilize the spin-flip producing a magnetic polaron which is further stabilized by lattice relaxation as in cuprates\cite{Anisimov1992,Yonemitsu1992,Dobry1994,Lorenzana1994}. We found stable polaron solutions  for both electron and hole doping of $\alpha$-AgF$_2$ as shown in Fig.~\ref{fig:polaron}(a) and (b). In both cases we find that the Ag polaron site becomes non-magnetic. 

\begin{figure}[t] 
 \centering
\includegraphics[width=9.5 cm]{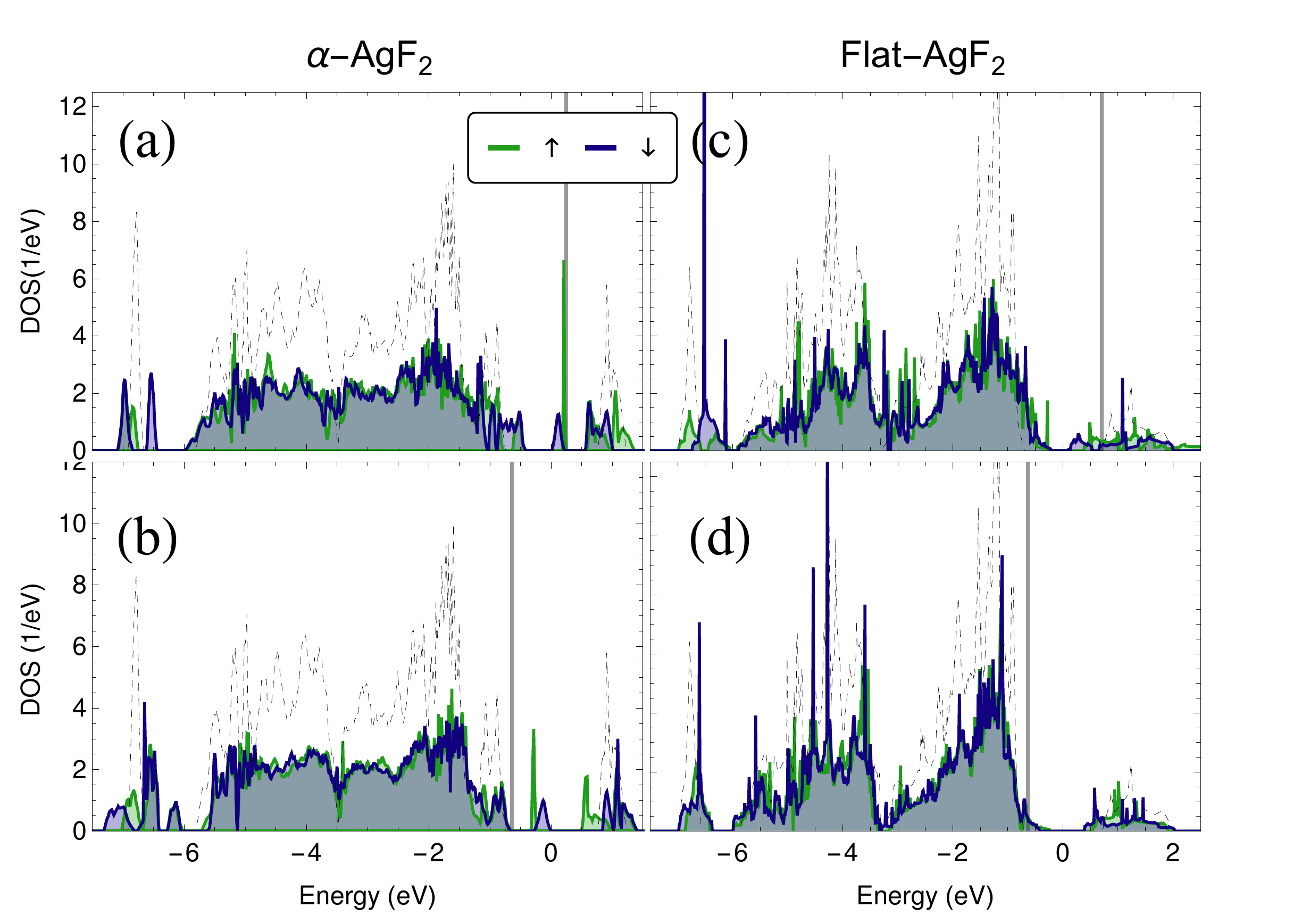} 
  \caption{Spin resolved total density of states for the various solutions in the same order as Fig.~\ref{fig:polaron}, namely electron  [(a) and (c)] and hole [(b) and  (d)] doping.  The zero of energy correspond to the middle of the gap of the undoped case. The undoped DOS is shown with a dashed line including both spin components. The doped case are shown approximately aligning the main bulk features with the undoped case.  The highest occupied state is indicated by the grey vertical line.
}
  \label{fig:dos}
\end{figure}

In the case of electron doping, Fig.~\ref{fig:polaron}(a), the polaron can be visualized as a local $d^{10}$ (Ag$^+$) state. Large relaxations of the lattice occur (red arrows) involving the F's nearest neighbors of the Ag$^+$ and also the second in-plane neighboring F's  producing a tightly bound self-trapped state. 
The bond with the nearest neighbor F's enlarges to 2.34\AA~(2.12\AA) 
while the Ag-F-Ag angle around the Ag$^+$ site decreases to  120$^\circ$ (129$^\circ$) where the values in parenthesis correspond to the undoped solution.  This strong self-trapping 
tendency can be understood considering that the undoped system has a quite 
narrow upper Hubbard band due to reduced $d$-$d$ hoping mediated by the ligand and leading to the narrow structure at $\approx 1$eV in the density of states [dashed line in Fig.~\ref{fig:dos}(a) and (b)]. As for superexchange, two $p$-orbitals in fluorine (one parallel and one perpendicular to the Ag-Ag bond) are relevant for the $d-d$ transfer which, however, interfere negatively \cite{Gawraczynski2019}.  Electron doping also tends to increase the buckling (reduced Ag-F-Ag angle) favoring the localization.   The situation is quite different in the case of hole doping [Fig.~\ref{fig:polaron}(b)]
 where lattice distortions of the nearest neighbor F's are 50\% smaller leading to an Ag-F distance of 2.00\AA~and an  Ag-F-Ag angle increased to 131$^\circ$. This leads locally to a flatter layer which opposes to the self-trapping. 

In cuprates, doped holes form Zhang-Rice singlets\cite{Eskes1988,Zhang1988} which, for large on-site Coulomb interactions, can be seen as a Heitler-London state between a $d_{x^2-y^2}$ hole and a hole in a $b_{1g}$ symmetrized combination of the surrounding $p$ orbitals. As the coupling with the lattice is increased, one can have a transition to a singlet molecular state\cite{Yonemitsu1992} with a quenched magnetic moment in the transition metal ion. This molecular singlet corresponds to the locally nonmagnetic solution found in hole doped $\alpha$-AgF$_2$ [Fig.~\ref{fig:polaron}(b)]. 
{Clearly, a Coulomb repulsion in the 4$d$'s orbitals smaller than in the 3$d$'s  contributes to favor such molecular singlet state over the Heitler-London state.} Notice that the hybridization matrix element\cite{Eskes1988} $T(b_{1g})$ between the $4d_{x^2-y^2}$ orbital and the combination of fluorine $p$ orbitals with the same symmetry is practically not affected by the buckling and favors the stability of  local singlet states. 

The different tendency between electrons and holes to form polarons is reversed in flat layers. Indeed, for the electron doped case a polaron is not stable in the sense that the found solution can be seen as a spin-flip with the added electron delocalized in the entire supercell. 
{The small displacements [barely visibly in  Fig.~\ref{fig:polaron}(c)]  are a magnetoelastic effect rather than polaronic.  }
To further check that a polaron is not stable we artificially increased the value of $U$ to 10 eV which leads to a well localized polaron solution with a sizable lattice distortion and a magnetic moment in the central Ag site reduced to 0.04 $\mu_B$. We then reduced $U$ (in steps of 1eV) to the physical value using the previous solution as a seed for the minimization and found that the added electron delocalizes gradually in all the cell with the magnetic moment reaching nearly  the same value in all Ag sites for $U=5$eV [Fig.~\ref{fig:polaron}(c)].
Since the polaron solution connects continuously with the localized solution we conclude that the polaron does not exist even as a metastable state in this case i.e. a bound state between the central spin flip and the added electron is not formed. 

For hole doping in the flat case, a polaron solution is found [Fig.~\ref{fig:polaron}(d)] {with a remnant magnetic moment in the central Ag site of 0.21$\mu_B$,} similar to the polaron found for cuprates in one of the first applications of the DFT+$U$ method\cite{Anisimov1992}. Notice that lattice distortions (red arrows) have been exaggerated in Fig.~\ref{fig:polaron} with different amplification factors for flat and buckled solutions.    

Figure~\ref{fig:dos} shows the total density of states (DOS) for up (green) and down spin (blue). In the case of $\alpha$-AgF$_2$, the locally nonmagnetic solutions reflect in two almost degenerate levels near the middle of the gap indicating two paired orbitals with very similar wave-functions but opposite spins.
One orbital is split off from the valence band and one from the conduction band. 
Both orbitals are occupied in the electron-polaron state leading to the non-magnetic Ag$^+$ site while both are empty in the hole doped case leading to the molecular orbital singlet state. 

For flat  AgF$_2$ [Fig.~\ref{fig:dos}(c),(d)], the ingap states are closer to the band edges consistent with shallow states or resonances. The $\downarrow$ ingap states in the electron doped case near 0.2 eV (c) corresponds to the spin flip in the polaron site [Fig.~\ref{fig:dos}(c)] which remains associated to a localized orbital.

\begin{figure}[t] 
 \centering
\includegraphics[width=8 cm]{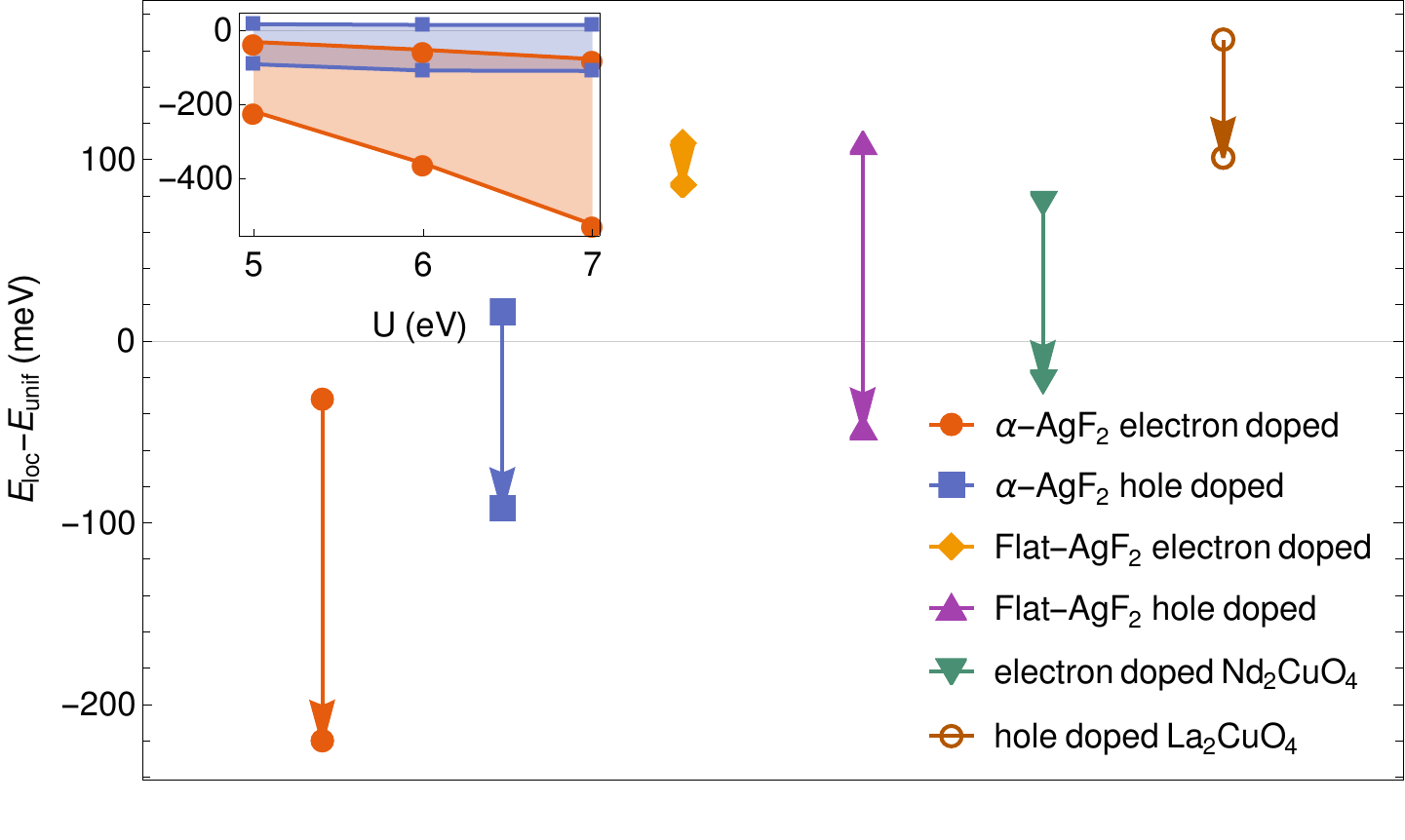} 
  \caption{Polaron binding energies for the systems studied. The upper point represents the energy gain by forming the polaron without lattice relaxation (a positive value indicates an energy cost). The lower point represents the extra gain obtained by relaxing the lattice.  The main panel is for $U=5$~eV while the inset shows the dependence of the results in $\alpha$-AgF$_2$ on the chosen value of $U$.}
  \label{fig:be}
\end{figure}

Figure \ref{fig:be} shows the binding energy of polarons defined as the total energy difference, $E_{loc}-E_{unif}$, between a polaronic solution and a uniform solution with the same number of electrons. The upper point is computed assuming a rigid lattice. A negative value indicates that the purely magnetic polaron is more stable than an uniformly doped state.   
The lower point is computed allowing the lattice to relax in the localized  solution, thus the length of the arrow represents the relaxation energy  $\epsilon_p$ associated with the formation of a phononic polaron on top of the magnetic polaron. We deem $\epsilon_p$ as the more important quantity to determine the possibility to achieve a metallic state. Indeed, the (positive or negative) binding energy in the rigid lattice corresponds to electronic degrees of freedom dressing the carrier. We know from $tJ$-model studies\cite{Dagotto1994}  that this leads to a magnetic polaron bandwidth which is approximately  $W\approx 2J$ which is still considerable and should not hamper metallicity. Weakening correlations compared to a $tJ$-model we expect the quasiparticle bandwidth to be even larger. $\epsilon_p$, instead represents relaxation of the much slower lattice degrees of freedom. For a characteristic phonon frequency $\omega_0$ the ratio  $\epsilon_p/\omega_0$ represents the number of quanta in the polaronic phonon cloud\cite{Mahan2000}. For large ratio and mapping to a Holstein small-polaron model\cite{Ranninger1992} one obtains an additional reduction of the bandwidth $W^*=W e^{-\epsilon_p/\omega_0}$ which may lead to a truly self-trapped state. Indeed, in the case of electron doped $\alpha$-AgF$_2$ we expect the magnetic polaron bandwidth to be $W\approx 140$ meV (using the experimental\cite{Gawraczynski2019} $J=70$~meV). Assuming an average phonon energy\cite{Gawraczynski2019,Tokar2020} of $\omega_0=50$~meV, one obtains an effective bandwidth $W^*$ of a few meV. This means that the electron is self-trapped and will behave as a classical particle unless the temperature is very low, in which case impurities will localize the polaron. 
The system is not expected to become metallic until very high levels of doping. 

For  hole doped $\alpha$-AgF$_2$ the situation is much better as the magnetic polaron is metastable (positive binding energy).  Coupling with the lattice creates a self-trapped state but $\epsilon_p$ is comparable to the 
expected $W\approx 2J=140 $ meV bandwidth. In this situation we expect that for small doping holes are self-trapped but moderate doping may lead to mobile carriers. Unfortunately, due to the very large work function\cite{Wegner2020} of AgF$_2$ such moderate hole doping may be hard to achieve. It is however encouraging that stoichiometric  AgF$_{2+x}$ compounds with $x=1/2,2/3,1$ are known to exist\cite{Zemva1991,Grochala2001}.

The prospect to achieve high-temperature superconductivity is much more improved in flat compounds. While hole doping still shows a large polaronic tendency (purple in Fig.~\ref{fig:be}) the more easily achievable electron doping is the one with less polaronic tendencies (orange). In this case, as discussed above, not even the magnetic polaron is stable and the positive binding energy reflects the magnetic energy associated with a spin flip in a uniformly doped system ($E_{loc}-E_{unif}\approx 2J$) accompanied with a very small lattice relaxation energy.

It is useful to compare the binding energies with similar results in cuprates.  For both types of doping we find a polaronic solution with the hole doped case being consistent with Ref.~\cite{Anisimov1992}. Interestingly, we find that the polaronic tendency is larger in electron-doped Nd$_{2}$CuO$_4$ than in hole-doped La$_2$CuO$_4$ as witnessed by a less positive binding energy of the magnetic polaron and a larger $\epsilon_p$ (respectively, green and brown lines in Fig.~\ref{fig:be}). This  may be the reason why antiferromagnetism persists up to a much higher doping ($\approx 0.15$) in the former than in the latter ($\approx 0.02$) and a similar asymmetry observed for the critical concentration to observe superconductivity\cite{Sobota2020}.  The similarity of $\epsilon_p$ in electron-doped Nd$_{2}$CuO$_4$ and hole doped AgF$_2$ suggests than metallization may be achieved in the latter if a similar high doping is achieved.  

One technical issue is the sensitivity of the present results to the chosen value of $U$ in the DFT+$U$ method.
As show in the inset of Fig.~\ref{fig:be}, the binding energy is practically independent of (increases with)  $U$ in the case of hole (electron) doped 
$\alpha$-AgF$_2$. Since the physical $U$ is probably\cite{Gawraczynski2019} larger than 5 eV our conclusions are not affected by this issue. For cuprates we find a similar trend, weak dependence for hole-doping and an increase of the binding energy with $U$ in the case of  electron doping which, again, does not change our conclusions.  

We have also considered an AgF$_2$ monolayer on top of a RbMgF$_3$ thick slab
 [Fig.~\ref{fig:structure}(c)], which was recently identified as an 
 optimum substrate to achieve flat layers\cite{Grzelak2020}. In this case we have not been able to stabilize polaronic solutions, neither in the electron nor in the hole doped case, but all attempts converged to uniform solutions in agreement with the above finding that flat layers are less favorable for polaron formation. In this case, enhanced rigidity of the lattice due to interaction with the substrate contributes to hamper the formation of polarons.

To conclude, we have studied the fate of doped carriers in AgF$_2$ with the prospect to achieve unconventional superconductivity. { In the electron doped $\alpha$-AgF$_2$ carriers are predicted to be self-trapped and unable to produce a metallic state. This suggests to concentrate efforts in the  
case of hole doping which is chemically more challenging but may lead to 
metallicity for a doping similar to electron doped cuprates ($\approx$ 0.15). An even more favorable situation is reached in recently predicted flat AgF$_2$\cite{Grzelak2020} on a suitable chosen substrate. In this case (or its three dimensional version) we find that a metallic state should be easily achievable.} By analogy with cuprates\cite{Grzelak2020} such state should support high-$T_c$ $d$-wave superconductivity at temperatures as large as nearly 200K.



\begin{acknowledgments}
The authors would like to thank J. Zannen for stimulating suggestions at the early stages of this work. J.L. acknowledges financial support from Italian MIUR through Project No. PRIN 2017Z8TS5B, and from Regione Lazio (L. R. 13/08) through project SIMAP. W.G. thanks the Polish National Science Center (NCN) for the Maestro project (2017/26/A/ST5/00570). This research was carried out with the support of the Interdisciplinary Centre for Mathematical and Computational Modelling (ICM), University of Warsaw under grant ADVANCE++ (no. GA76-19).
M.D. acknowledges the European Regional Development Fund, Research and Innovation Operational Program (project No. ITMS2014+: 313011W085), the Slovak Research and Development Agency (grant No. APVV-18-0168) and Scientific Grant Agency of the Slovak Republic (grant No. VG 1/0223/19). 
\end{acknowledgments}


\end{document}